\documentclass[%
 reprint,
 superscriptaddress,
 amsmath,amssymb,showpacs,
 aps,
 prb,
]{revtex4-1}

\usepackage{graphicx}
\usepackage{dcolumn}
\usepackage{bm}
\usepackage{mathrsfs}
\usepackage{float}
\usepackage{color}

\begin{document}

\preprint{APS/123-QED}

\title{First-principles embedded-cluster calculations\\ of the neutral and charged oxygen vacancy at the rutile TiO$_2$(110) surface}

\author{Daniel Berger}
 \email{daniel.berger@ch.tum.de}
 \affiliation{Chair for Theoretical Chemistry and Catalysis Research Center, Technische Universit{\"a}t M{\"u}nchen,\\ Lichtenbergstr. 4, D-85747 Garching, Germany}
\author{Harald Oberhofer}
 \affiliation{Chair for Theoretical Chemistry and Catalysis Research Center, Technische Universit{\"a}t M{\"u}nchen,\\ Lichtenbergstr. 4, D-85747 Garching, Germany}
\author{Karsten Reuter}
\affiliation{Chair for Theoretical Chemistry and Catalysis Research Center, Technische Universit{\"a}t M{\"u}nchen,\\ Lichtenbergstr. 4, D-85747 Garching, Germany}
\affiliation{SUNCAT Center for Interface Science and Catalysis, SLAC National Accelerator Laboratory \& Stanford University,
443 Via Ortega, Stanford, CA 94035-4300, U.S.A.}

\date{\today}

\begin{abstract}
We perform full-potential screened-hybrid density-functional theory (DFT) calculations to compare the thermodynamic stability of neutral and charged states of the surface oxygen vacancy at the rutile TiO$_2$(110) surface. Solid-state (QM/MM) embedded-cluster calculations are employed to account for the strong TiO$_2$ polarization response to the charged defect states. Similar to the situation for the bulk O vacancy, the +2 charge state $V_{\rm O}^{2+}$ is found to be energetically by far most stable. Only for Fermi-level positions very close to the conduction band, small polarons may at best be trapped by the charged vacancy. The large decrease of the $V_{\rm O}^{2+}$ formation energy with decreasing Fermi-level position indicates strongly enhanced surface O vacancy concentrations for $p$-doped samples.
\end{abstract}


\maketitle

\section{Introduction}

Its seemingly endless range of applications \cite{fujishima1972photolysis,graezel_photoelectro,graezelcell,TiO2_data_storage,chen2007titanium}
has made TiO$_2$ one of the most studied transition metal oxides to date. Much of this material's functionality in corresponding (opto-)electronic, (photo-)catalytic or photovoltaic applications derives not from the ideal bulk and surface structures, but is instead critically determined by intrinsic defects.\cite{jupille2015defects}
Among these, the oxygen vacancy and in particular its nature as a charge trapping center have been most controversially discussed.\cite{minato2014atomic} The removal of an O atom from the bulk TiO$_2$ lattice results in a single-particle defect state created from the three Ti dangling bonds that point into the vacancy. The energetic position of the state depends sensitively on its electron occupancy and concomitant local lattice relaxations. This occupancy can range from two electrons in the charge-neutral defect state (V$^0_{\rm O}$), over one electron in the singly-charged defect state (V$^{+}_{\rm O}$), to empty in the doubly-charged state (V$^{2+}_{\rm O}$). Formerly prevalent seemingly contradictory schools of thought viewed the vacancy either as a shallow donor (V$^{2+}_{\rm O}$) that contributes to the $n$-type conductivity, \cite{PhysRevB.54.7945,doi:10.1021/jp0606210,PhysRevB.48.12406} or as an electrically inactive deep trap (V$^0_{\rm O}$) that reduces neighboring Ti atoms and thus e.g. rationalizes an experimentally observed gap state about 0.8\,eV below the conduction band.\cite{doi:10.1080/00018730110103249,0953-8984-19-25-255208,Eagles19641243,PSSB:PSSB19680270144}

For bulk rutile TiO$_2$ hybrid-functional density-functional theory (DFT) calculations by Janotti {\em et al.} \cite{janotti2010hybrid,PSSR:PSSR201206464} recently resolved the preceding discrepancies by showing that $V^{2+}_{\rm O}$ is the thermodynamically by far most stable configuration for all Fermi-level positions within the band gap. However, this charged donor state can trap one or two small polarons, in which excess electrons are localized on neighboring Ti$^{3+}$ sites. The resulting weakly bound complex of shallow donor and small polarons then naturally explains all experimental findings, but has to be carefully distinguished from the neutral V$^0_{\rm O}$ configuration in which the electrons occupy the defect state centered on the O vacancy site itself.

As it is particularly surface O vacancies that play a crucial role in many of the TiO$_2$ material applications \cite{Diebold2003,doi:10.1021/cr00035a013,GandugliaPirovano2007219,doi:10.1021/ja207826q}, it is important to assess how much of this novel understanding derived for the bulk transfers also to these surface defects. Corresponding first-principles calculations are, however, rather demanding. Already 
in the bulk case, at least hybrid-functional DFT is required to achieve an appropriate electron localization.~\cite{Selloni2006,janotti2010hybrid} Even with present-day computing power this constrains the system sizes that can be accessed. At the same time, the huge dielectric constant of TiO$_2$ leads to a very large dielectric response in case of the charged defects. Here, not only lattice relaxations in the direct vicinity of the defect but the polarization of the entire semi-infinite surrounding medium contribute significantly. Within the conventional periodic boundary condition (PBC) supercell approach this requires intricate extrapolation procedures involving supercells of increasing size.\cite{makov1995periodic,PhysRevLett.110.095505,RevModPhys_86_253}

In this situation we instead opt for first-principles embedded-cluster calculations, in which the employed full-potential scheme allows for a numerically particularly efficient application of hybrid-functional DFT inside the quantum mechanic (QM) cluster region.~\cite{me} In the extended molecular mechanic (MM) embedding region appropriately optimized interatomic potentials provide a quantitative account of the strong TiO$_2$ polarization response. We use this setup specifically to compute the formation energies, structural relaxations, and electronic structure of the bridging O vacancy at the rutile TiO$_2$(110) surface. Consistent with the bulk calculations of Janotti {\em et al.} we find that over a wide range of Fermi-level positions and oxygen chemical potentials the doubly-charged V$^{2+}_{\rm O}$ state is thermodynamically clearly favored. The steeply decreasing formation energy of this charged defect with a lowering Fermi level then suggests $p$-doping as a promising avenue to tune the surface vacancy concentrations and therewith the catalytic activity of this important material.

\section{Methodology}

\begin{figure}[H]
\centering
\includegraphics[width=8.5 cm]{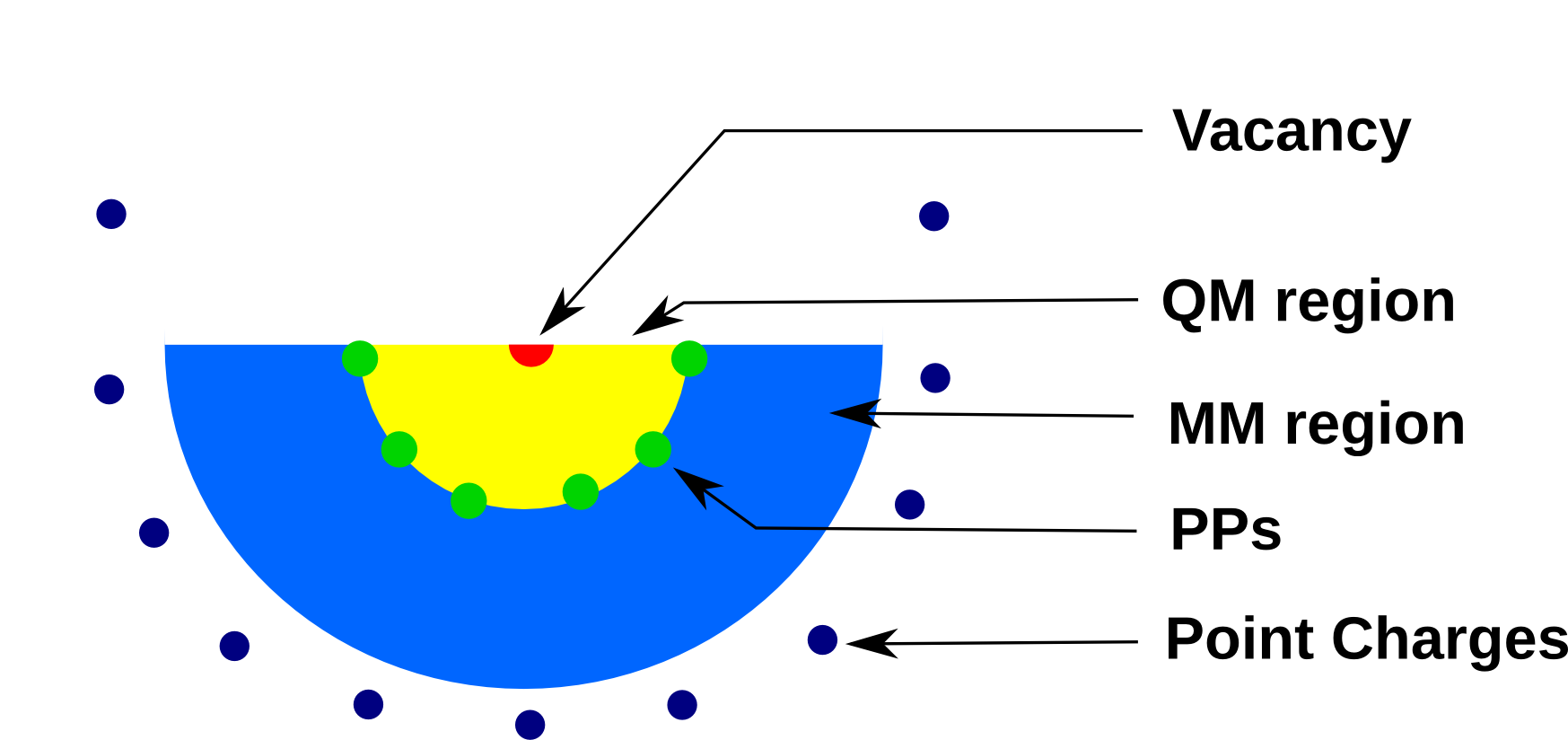}
\caption{Schematic representation of the employed concentric solid-state embedding approach. A quantum-mechanical (QM) region hosting the surface oxygen vacancy is surrounded by a molecular mechanics (MM) region represented by a polarizable interatomic potential. Spurious charge leakage out of the QM region is prevented through a transition shell in which cations are described with pseudopotentials (PPs). The full electrostatic potential of the infinite crystal surface is reproduced by placing point charges with fitted values around the MM region.}
\label{fig1}
\end{figure}

We describe the localized surface defect in a concentric solid-state embedding approach as sketched in Fig. \ref{fig1}. The immediate vicinity of the defect and the rehybridization induced through it is treated at the QM level, and in particular through DFT. This QM region is embedded into a much larger MM region, which accounts for the longer-ranged dielectric properties of the TiO$_2$(110) surface on the level of a polarizable interatomic potential. In a transition shell at the QM/MM boundary, cations are described by pseudopotentials (PPs) to prevent a spurious overpolarization of the electron density (aka charge leakage). These PPs recover the long-range electrostatics of a point charge, but also have a repulsive short-range contribution which effectively mimics Pauli repulsion of core electrons. A final exterior shell of point charges is added at the outer boundary of the finite MM region. These point charges have values that are fitted to reproduce the full electrostratic potential of the infinite surface inside the QM region.\cite{sokol2004hybrid} The next consecutive sections provide details of the DFT calculations, the parametrization of the interatomic potential, and the employed surface models within this overall approach.

\subsection{Density-functional theory calculations}

All DFT calculations have been performed with the full-potential, all-electron framework {\tt FHI-aims} \cite{blum2009ab,ren2012resolution}. 
Electronic exchange and correlation (xc) is treated at the level of screened hybrid-DFT, applying the HSE06 functional with the default mixing of 25\% exact exchange and the default screening parameter of 0.2 \AA$^{-1}$.\cite{HSE06} Systematic test calculations showed that the \textit{tier2} numerical atomic orbital basis set and the default \textit{tight} settings for the atom-centered integration grids ensure
a numerical convergence of the calculated defect formation energies within $\pm 10$\,meV. Spin polarization is included throughout. As PPs for the transition shell we employ {\tt FHI98PP}-generated \cite{fuchs1999ab} Ti$^{4+}$ Kleinman-Bylander PPs \cite{kleinman1982efficacious} with non-local projector functions expanded up to the $d$-states. Further details of these potentials and their implementation into
{\tt FHI-aims} can be found in ref. \onlinecite{me}.

\begin{table}
\begin{tabular}{l|cc|cc|cc}
\hline
      & $a$ [in {\AA}] & $c$ [in {\AA}] & $\epsilon^a_\circ$ & $\epsilon^c_\circ$ & $\epsilon^a_\infty$ & $\epsilon^c_\infty$ \\ \hline
Exp. \cite{burdett1987structural,traylor1971lattice,parker1961static}
                                           & 4.587 & 2.954 & 111  & 257  & 6.84 & 8.43 \\ \hline
HSE06 \cite{lee2011influence}              & 4.588 & 2.951 & 278  & 402  & 5.74 & 6.77 \\
HSE06 (this work)                          & 4.588 & 2.951 &      &      &      &      \\ \hline
MM (this work)                             & 4.587 & 2.950 & 3    & 12   & 5.76 & 6.73 \\ \hline
\end{tabular}
\caption{Rutile TiO$_2$ lattice constants $a$ and $c$, as well as its static ($\epsilon^a_\circ$, $\epsilon^c_\circ$) and high-frequency ($\epsilon^a_\infty$ and $\epsilon^c_\infty$) dielectric constants along the corresponding axes. Literature data from experiment \cite{burdett1987structural,traylor1971lattice,parker1961static} and DFT-HSE06 calculations \cite{lee2011influence} are compared against our own calculations at the DFT-HSE06 level and with the parametrized interatomic potential for the MM region (see text).} 
\label{table1}
\end{table}

The finite clusters describing the QM region are constructed using optimized bulk lattice parameters and relaxed positions of surface atoms at the ideal TiO$_2$(110) surface as obtained from PBC-DFT supercell calculations. The calculated lattice parameters of rutile TiO$_2$ are listed in Table \ref{table1} and agree very well with experiment \cite{burdett1987structural} and preceding calculations at HSE06 level \cite{lee2011influence}. The surface calculations employed symmetric 5 O-Ti$_2$O$_2$-O trilayer slabs, a $c(4 \times 2)$ surface unit-cell, and $( 4 \times 4 \times 1)$ Monkhorst-Pack k-point sampling \cite{MonkhorstPack}. Periodic slabs were separated in z-direction by 40 {\AA} of vacuum and electronically decoupled through a dipole correction. The surfaces were fully relaxed until residual forces fell below 10$^{-3}$ eV/{\AA}.

Within PBCs the zero reference of the electrostatic potential is not uniquely defined.\cite{saunders1992electrostatic} Notwithstanding, in surface calculations, the vacuum level can be determined as the electrostatic potential at the middle of the vacuum separating the slabs. This gives access to the work function and the absolute position of the valence band maximum (VBM) at the surface, $\epsilon_{\rm VBM(surf)} = -8.2$\,eV. To also access the bulk VBM position we computed the layer-resolved Ti$_{1s}$ core-level positions in an 11-trilayer thick $(1 \times 1)$ slab. These positions indicate a negligible band bending of the order of 30\,meV for the ideal stoichiometric TiO$_2$(110) surface. This finding is confirmed by additional calculations with up to 27-trilayer thick slabs using the PBE xc functional \cite{PBE1996}. For the purposes of this work we therefore equate surface and bulk VBM levels, and henceforth refer only to $\epsilon_{\rm VBM}$.

\subsection{Parametrization of the interatomic potential}

The HSE06 functional achieves a reliable account of the electronic contribution to the bulk dielectric properties of rutile TiO$_2$, as represented by the close match of the high-frequency dielectric constants $\epsilon^a_\infty$ and $\epsilon^c_\infty$ in Table \ref{table1}. In contrast, it fails largely to describe the dominant lattice contribution additionally contained in the large static dielectric constants
$\epsilon^a_\circ$ and $\epsilon^c_\circ$. This has been traced back to its deficiencies in describing the intricate soft phonon modes of this material.\cite{lee2011influence} In order to achieve a seamless embedding the employed interatomic potentials in the MM region should generally match the dielectric properties of the xc functional employed in the QM region. In the present case this would mean that the QM/MM approach then exhibits the same shortcomings with respect to the static dielectric properties; a point to which we will return in the discussion part below. In addition to the dielectric properties, the MM potential also has to match the QM lattice constants, to avoid artificially confining stress in particular when geometric relaxation of the QM region is to be considered. These demands highly challenge any existing interatomic potential.\cite{catlow1988interatomic,collins1996molecular} In the present case, this situation is further aggravated by the necessity to saturate the QM region with cationic norm-conserving PPs, which by definition have integer charges; in the case of Ti a charge of +4. For consistency, the remaining MM region is then also restricted to formal charges $+4$ and $-2$ on Ti cations and O anions, respectively.

Oxygen ions in TiO$_2$ are highly polarizable, and are in fact intrinsically polarized in the rutile structure. Using as interatomic potential a simple rigid ion model with formal charges does not capture this physics. In the QM/MM context, such potentials lead to an overpolarization at the QM cluster region boundary and an overestimation of the electrostatic potential.\cite{reinhardt1996adsorption,sanz2000first} In contrast, oxygen polarizability can be modeled efficiently within a polarizable shell-model \cite{dick1958theory}, as has recently been demonstrated for TiO$_2$ by Scanlon {\em et al.} \cite{scanlon2013band}. Here, the oxygen anion is described by two point charges: A ``core'' charge representing the nuclei and closed-shell core electrons, and a ``shell'' charge simulating the valence electron cloud. Mimicking electronic polarizability the oxygen core (c) and its shell (s) interact via a spring potential
\begin{equation}
V_{\rm c-s} \;=\; k_{\rm c-s} \; r^2_{\rm c-s} \left[ {\rm cosh}\left( \frac{d_{\rm c-s}}{r_{\rm c-s}} \right) - 1 \right] \quad ,
\end{equation}
where $d_{\rm c-s}$ is the distance between the core and shell charge, and $k_{\rm c-s}$ and $r_{\rm c-s}$ are parameters defining the potential. In this model \cite{dick1958theory}, the dominant Coulomb interaction between different oxygen shells (s-s) and between oxygen shells and Ti point charges (s-Ti) is furthermore augmented by Buckingham potentials
\begin{equation}
V_{\rm s-s} \;=\; A_{\rm s-s} \; {\rm exp}\left(- \frac{d_{\rm s-s}}{\rho_{\rm s-s}}\right) - \frac{C_{\rm s-s}}{d^{6}_{\rm s-s}} 
\end{equation}
and
\begin{equation}
V_{\rm s-Ti} \;=\; A_{\rm s-Ti} \; {\rm exp}\left(- \frac{d_{\rm s-Ti}}{\rho_{\rm s-Ti}}\right) - \frac{C_{\rm s-Ti}}{d^{6}_{\rm s-Ti}} \quad ,
\end{equation}
to provide an effective account of dispersive interactions and Pauli repulsion. Here, $d_{\rm s-s}$ and $d_{\rm s-Ti}$ are the distances between oxygen shells and between oxygen shell and Ti, respectively, and $A$, $\rho$ and $C$ are potential parameters. Restricting the interaction between MM Ti cations to the mere formal charge electrostatics, the model is thus defined through a set of nine parameters: $[ k_{\rm c-s}, r_{\rm c-s}, A_{\rm s-s}, \rho_{\rm s-s}, C_{\rm s-s}, A_{\rm s-Ti}, \rho_{\rm s-Ti}, C_{\rm s-Ti}, q_{\rm s} ]$, with a final parameter $q_{\rm s}$ for the charge on the oxygen shell.

To obtain a seamless match to the QM region we perform a global optimization of these parameters to represent the bulk DFT-HSE06 lattice and dielectric constants. Specifically, we employ a differential evolutionary algorithm \cite{storn1997differential} from the Python package {\tt Inspyred 1.0}\cite{inspyred} to minimize the dimensionless cost function
\begin{equation}
F \;=\;  \sqrt{ \sum_i \left( \frac{L_i^{\rm MM} - L_i^{\rm DFT-HSE06}}{L_i^{\rm DFT-HSE06}}\right)^2 } \quad ,
\end{equation}
with $L_i = [a,c,\epsilon^a_\circ,\epsilon^c_\circ, \epsilon^a_\infty, \epsilon^c_\infty ]$ and using the DFT-HSE06 values from ref. \onlinecite{lee2011influence}, cf.~Table \ref{table1}, as target values. In the corresponding MM calculations, the internal lattice parameter $u$ is always kept at its DFT-HSE06 value (0.305), and the static and high-frequency dielectric tensors are determined from the second derivative matrix of all MM particles or only of the shells, respectively.\cite{gale1997gulp}

\begin{table} [h]
\centering
\begin{tabular}{lccc}
\hline
         & $k_{\rm c-s}$ [in eV/{\AA}$^{-2}$] & $r_{\rm c-s}$ [in {\AA}] & $q_{\rm s}$ [in e] \\ \hline
         & 23.67                 & 0.098                 & -2.9332             \\ \hline \hline
				 & $A$ [in eV]           & $\rho$ [in {\AA}]        & $C$ [in eV {\AA}$^{6}$]         \\ \hline
s - s    & 23550                 & 0.2113                & 38.55               \\
s - Ti   & 1838                  & 0.3207                & 26.62               \\ \hline

\end{tabular}
\caption{Interatomic potential parameters, optimized to reproduce the bulk TiO$_2$ DFT-HSE06 lattice parameters and high-frequency dielectric constants, see text.}
\label{table2}
\end{table}

The best parameter sets generated this way still exhibit rather large errors, with $F \ge 0.33$. They typically exhibit substantial deviations in the lattice constants. This agrees not only with the observation from Catlow {\em et al.}\cite{catlow1988interatomic} who assigned the inability to reproduce both lattice and static dielectric constants to missing many-body terms in this class of interatomic potentials. It is also consistent with the finding of Lee {\em et al.} \cite{lee2011influence} that deficiencies in the description of the TiO$_2$ soft phonon modes (and therewith static dielectric constants) can be effectively cured through the use of different lattice constants. In the present context, an accurate representation of the lattice constants is indispensable though, to avoid artificial stress on the QM region. We therefore removed the static dielectric constants from the target set and immediately obtain parameter sets with significantly reduced cost functions. The best parameter set exhibits an $F = 0.007$ and is used in all QM/MM calculations reported below. It is compiled in Table \ref{table2} and yields highly accurate lattice parameters and high-frequency dielectric constants as shown in Table \ref{table1}. Notwithstanding, its largely erroneous representation of the static dielectric properties (which is actually of the same magnitude but opposite sign to those obtained with HSE06 itself) is a concern and we discuss in Section III how this is addressed in our defect calculations.

\subsection{QM/MM setup}

\begin{figure}[!]
\centering
\includegraphics[width=6.5 cm]{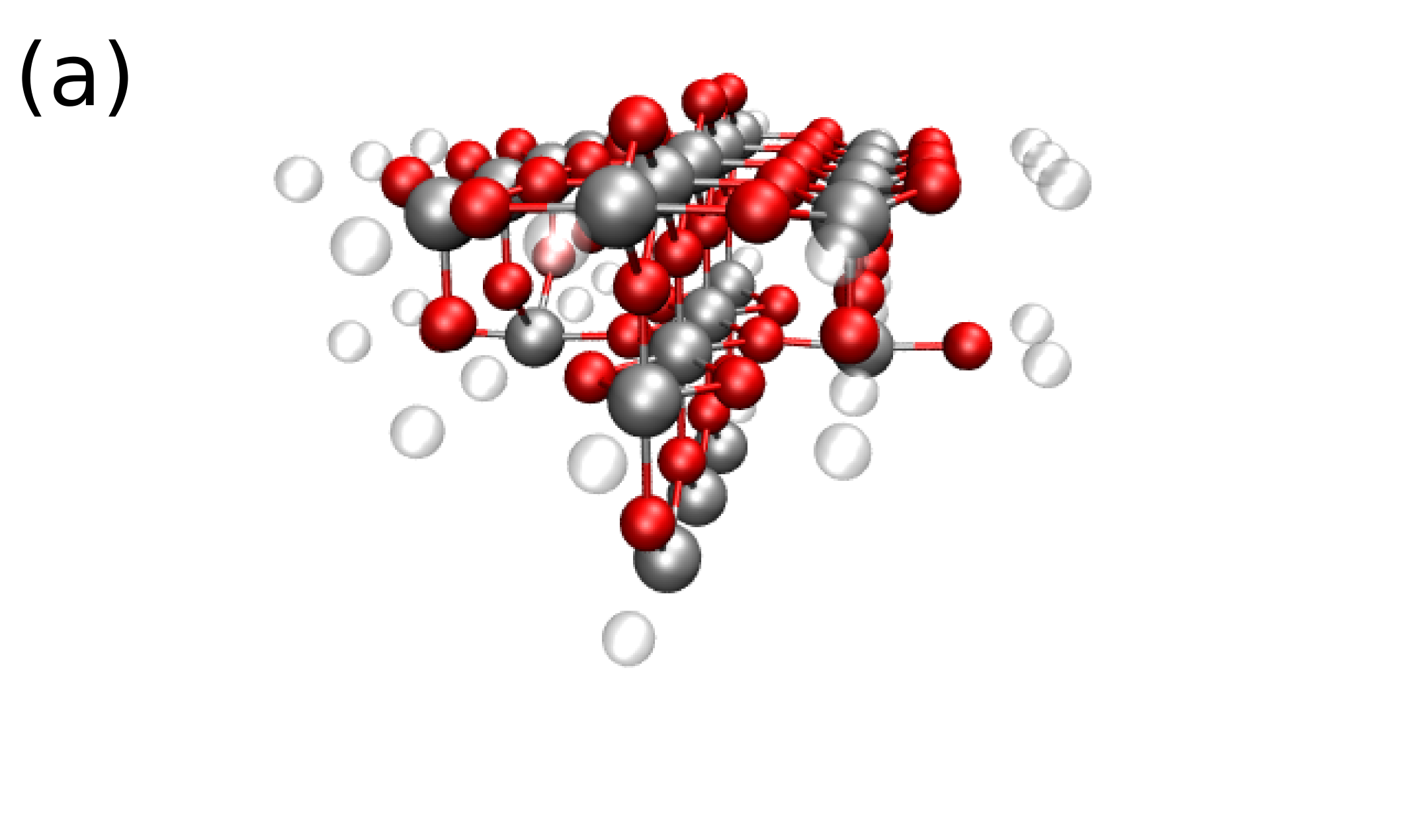}
\includegraphics[width=6.5 cm]{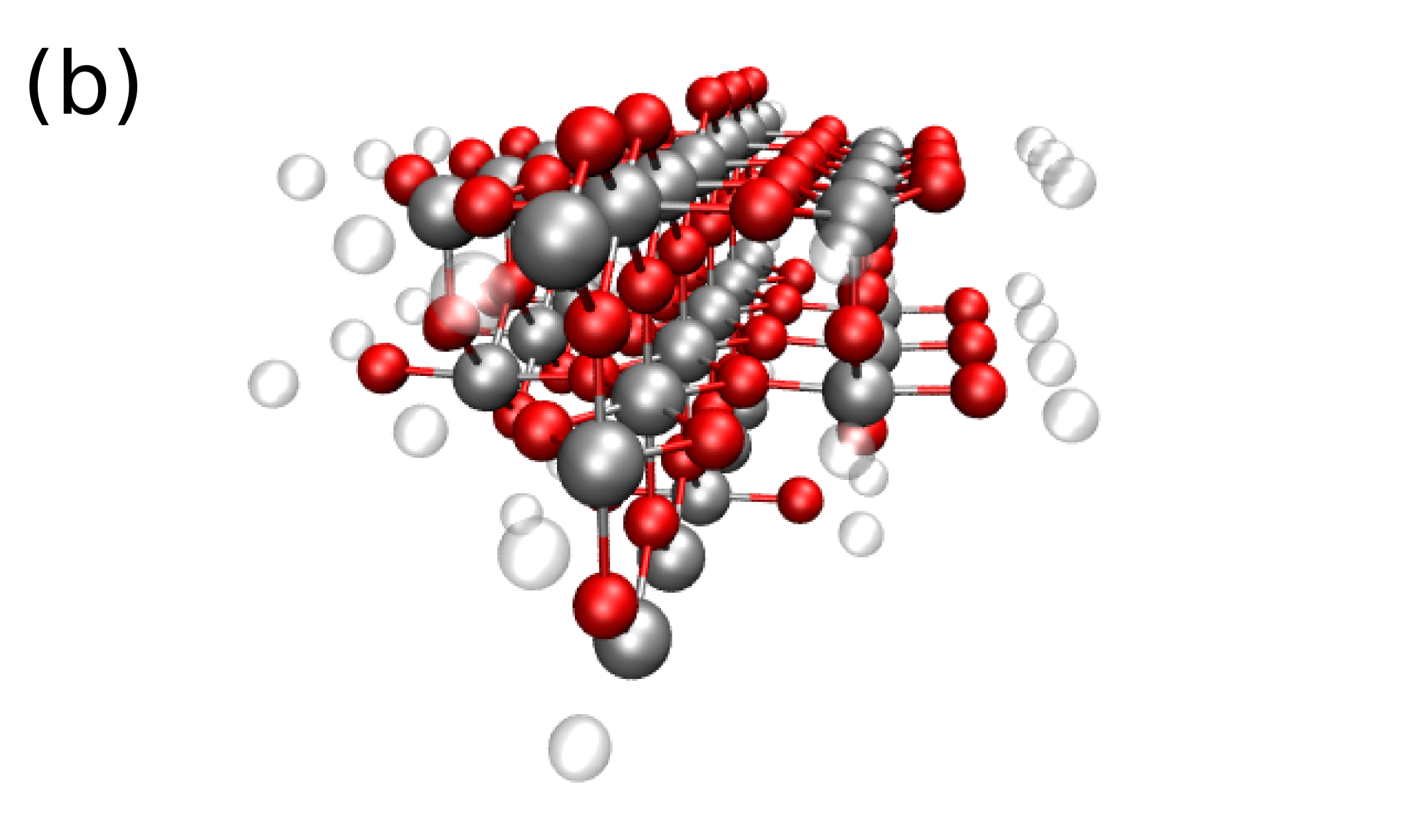}
\includegraphics[width=6.5 cm]{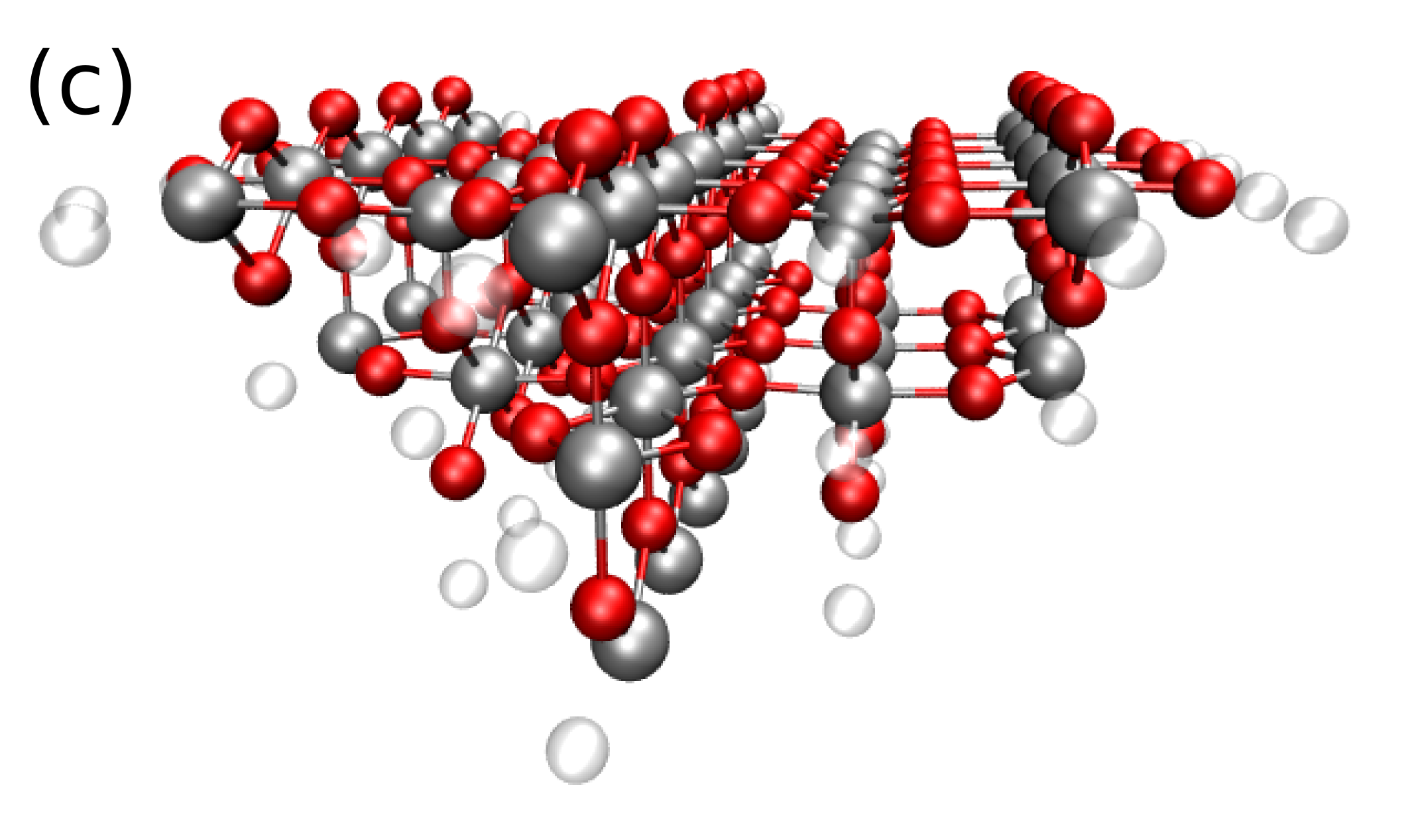}
\caption{Perspective view of the employed (a) Ti$_{22}$O$_{43}$, (b) Ti$_{32}$O$_{63}$ and (c) Ti$_{46}$O$_{91}$ clusters, each exhibiting a surface O vacancy in their central bridging O atom row. Ti atoms are shown as large white spheres, O atoms as small red spheres, and semi-transparent gray spheres mark the positions where PPs represent the immediately surrounding Ti-cations of the MM region.}
\label{fig2}
\end{figure}

In order to assess the convergence with respect to the employed QM region all calculations are done for a sequence of three embedded cluster models, originally suggested by Ammal and Heyden \cite{Heyden2010}. As shown in Fig.~\ref{fig2}, all three clusters, Ti$_{22}$O$_{43}$, Ti$_{32}$O$_{63}$ and Ti$_{46}$O$_{91}$, are centered on a bridging oxygen row, from where the central O atom has been removed to create the
surface vacancy. Only the smallest and the largest cluster are of approximately hemispherical shape, while the middle Ti$_{32}$O$_{63}$ cluster is rather hemi-ellipsoidal, owing to the specifics of the rutile structure. Each cluster is embedded into an MM region that extends hemispherically up to a constant outer radius of 25\,{\AA}. For the smallest cluster this translates to a total number of 3029 MM atoms, while for the largest cluster this translates to 2957 MM atoms. Every MM cation within 6\,{\AA} vicinity of the QM cluster is replaced by PPs to suppress spurious charge leakage out of the QM region, cf.~Fig.~\ref{fig2}. All MM cores are placed at positions according to those of the fully relaxed DFT supercell reference calculation for the stoichiometric TiO$_2$(110) surface. All O shells are initialized to the fully relaxed state within a corresponding MM supercell reference calculation with the MM cores at exactly the same positions. A final exterior shell of 64 point charges around the MM region is then added, with charges fitted to reproduce the full electrostatic embedding potential of an infinite TiO$_2$(110) surface.~\cite{sokol2004hybrid} Finally, for every QM cluster all MM shells and QM atom positions are fully relaxed. This setup defines what will henceforth be referred to as the ideal TiO$_2$(110) surface. The surface defect setups are created from this reference setup, either without subsequent geometry relaxation ("non-relaxed" geometries) or with full geometry relaxation of all QM atoms except those at the QM region boundary until residual forces are below 10$^{-3}$ eV/{\AA} ("relaxed" geometries).

Test calculations increasing the outer radius of the MM region up to 30\,{\AA} show full convergence of the formation energy and electronic structure in the case of the neutral defect. In case of a net-charged QM region the polarization response is much longer ranged though ($\propto q/(\epsilon R)$). Fortunately, the missing polarization energy outside of the finite MM region can be reliably captured through an analytical correction. For a hemisphere with radius $R$ in a continuum with a dielectric constant $\epsilon$ and carrying a charge $q$ at its center this analytic correction can be derived as \cite{sokol2004hybrid}
\begin{equation}
\label{eq:pol_surface}
\Delta E_{\rm pol}(q) \;=\; - \frac{q^2}{2R} \frac{\epsilon - 1}{\epsilon + 1} \quad .
\end{equation}
Using the high-frequency dielectric constant in this expression as further discussed below, we validated that with this post-correction also the formation energies of charged defects are fully converged with respect to the size of the MM region. Technically, we hereby use as isotropic dielectric constant the average over the diagonal entries of the bulk dielectric tensor, $\epsilon = (2\epsilon^a_\infty + \epsilon^c_\infty)/3$. As radius $R$ we simply set the outer radius of the MM region and variations of $R$ by $\pm 1$\,{\AA} have a negligible effect on the calculated formation energies. As an approximate method we also use the analytical correction directly outside the QM region ({\em vide infra}). Here, the choice of the radius is more critical and we determine it by measuring the semi-principal axes of the hemi-ellipsoid defined by the atomic positions of the QM cluster. $R$ is then taken as the radius of a hemisphere with identical volume. Uncertainties of $\pm 5$\,\% in the thus determined radius translate in this case into an uncertainty of $\pm 0.2$\,eV for $\Delta E^{\rm f}({\rm V}^{2+}_{\rm O})$ and of $\pm 0.05$\,eV for $\Delta E^{\rm f}({\rm V}^{+}_{\rm O})$.

The actual QM/MM calculations for the thus defined setup are performed within the {\tt ChemShell} environment \cite{sherwood2003quasi,sokol2004hybrid} with the interface described in detail before.~\cite{me} We specifically use {\tt GULP} \cite{gale1997gulp} for the MM force calculations and the {\tt DL-FIND} routine \cite{kaestner2009dl} for the geometry optimizations. Self-consistent polarization, aka shell optimization, within the MM region as a response to an updated QM geometry is hereby calculated in a series of micro-iterations.

\subsection{Defect formation energies}
\label{subsection:defects}

Neglecting vibrational entropic contributions in the solid state and defect-defect interactions in the dilute limit we define the formation energy to create a surface O vacancy in charge state $q$ as\cite{walle2004}
\begin{eqnarray} 
\Delta E^{\rm f}({\rm V}^q_{\rm O}) &=&  E({\rm V}^q_{\rm O}) - E({\rm TiO_2(110)})  \;+ \nonumber \\ 
&+& \mu_{\rm O} + q \epsilon_{\rm F} + \Delta E_{\rm pol}(q)  \quad .
\label{formation}
\end{eqnarray}
Here, $E({\rm TiO_2(110)})$ and $E({\rm V}^q_{\rm O})$ are the total energies of ideal TiO$_2$(110) and of TiO$_2$(110) with the defect as obtained from our QM/MM calculations, respectively. $\mu_{\rm O}$ is the chemical potential of oxygen and $\epsilon_{\rm F}$ is the Fermi energy. In our sign convention a positive formation energy implies a cost to create the defect. Correspondingly, the charge state exhibiting the lowest formation energy will be the thermodynamically stable state.

$\mu_{\rm O}$ represents the energy of the reservoir which takes up the O atom that is removed from the crystal. It is generally a variable. If the reservoir is a surrounding oxygen gas phase, $\mu_{\rm O}$ is e.g. dependent on temperature and O$_2$ pressure. Limits for $\mu_{\rm O}$ can, however, be derived \cite{PhysRevB.65.035406}. In the extreme O-rich limit
\begin{equation}
\mu_{\rm O}({\rm O-rich}) \;=\; 1/2 E({\rm O}_2) \quad ,
\end{equation}
with $E({\rm O}_2)$ the total energy of an isolated O$_2$ molecule. The opposite O-poor (Ti-rich) limit can be assessed from the stability condition of bulk TiO$_2$ against decomposition into Ti$_2$O$_3$, $\mu_{\rm O} > - \Delta E^{\rm f}({\rm Ti_2O_3}) + 2\Delta E^{\rm f}({\rm TiO_2})$, where $\Delta E^{\rm f}({\rm Ti_2O_3})$ and $\Delta E^{\rm f}({\rm TiO_2})$ are the formation energies of bulk Ti$_2$O$_3$ and TiO$_2$, respectively. At the HSE06 level this yields \cite{janotti2010hybrid}
\begin{equation}
\mu_{\rm O}({\rm O-poor}) \;=\; \mu_{\rm O}({\rm O-rich}) \;-\; 4.07\,{\rm eV} \quad .
\label{O-poor}
\end{equation}

The Fermi energy $\epsilon_{\rm F}$ is the energy of the reservoir where electrons released from the charged defects move to. Similar to $\mu_{\rm O}$, also $\epsilon_{\rm F}$ is a variable that can e.g.~be tuned through doping. It is most conveniently referenced with respect to $\epsilon_{\rm VBM}$ and below we thus report values from zero (Fermi level positioned at the VBM) to $\Delta \epsilon_{\rm gap}$ (Fermi level positioned at the conduction band minimum (CBM)), with $\Delta \epsilon_{\rm gap}$ the bulk band gap. In our QM/MM setup the zero reference for the Madelung potential and hence the VBM is by construction the vacuum level. In order to compute $\Delta E^{\rm f}$ through eq.~(\ref{formation}) the offset of the absolute $\epsilon_{\rm F}$ value by $\epsilon_{\rm VBM}$ must therefore be considered.

Within our supercell setup we calculate a $\Delta \epsilon_{\rm gap} = 3.15$\,eV, which agrees well with preceding studies at the HSE06 level (3.19\,eV \cite{0953-8984-24-43-435504}, 3.31\,eV \cite{CCTC:CCTC201100498}, 3.05\,eV \cite{janotti2010hybrid}, with the last study only employing 20\% exact exchange). This also extends to the previously calculated absolute VBM position\cite{doi:10.1021/ct500087v}, which we determine at $-8.2$\,eV. The latter value is also consistent with the experimental value which can be estimated from the experimental work function and a presumed Fermi level position about 0.1-0.2\,eV below the CBM \cite{PhysRevB.71.235416} in corresponding samples \cite{schierbaum1996interaction,PhysRevB.70.045415,PhysRevB.71.235416}.

In an ideal world, $\epsilon_{\rm VBM}$ calculated for the different QM cluster sizes within our QM/MM setup would always be identical to the value obtained with the PBC supercell calculations. In practice, an imperfect QM/MM coupling affects the electronic structure and there in particular those orbitals with appreciable overlap with the QM/MM boundary. In the present case, this concerns precisely the highest occupied molecular orbital (HOMO), a.k.a.~the VBM, of the QM clusters, which is of a rather delocalized nature. This results in a slight spurious electron depletion or aggregation in the QM region depending on the shape and size of the QM cluster, and therewith to shifts of the overall electrostatic potential. In order to compensate for this shift we apply a further correction term in the calculation of the absolute Fermi level position for eq.~(\ref{formation}). This correction is determined once for each cluster size as the difference in the calculated unrelaxed, closed-shell singlet V$^0_{\rm O}$ defect-level position as compared to the corresponding level in the PBC calculation. In contrast to the VBM/HOMO, the defect level is well localized in the center of the cluster. This minimizes short-range artifacts from the QM/MM interface and probes (and corrects) the electrostatic potential exactly in the relevant region.

\section{Results \& Discussion}

\subsection{Electronic defect structure}

\begin{figure}
\centering
\includegraphics[width=8.5 cm]{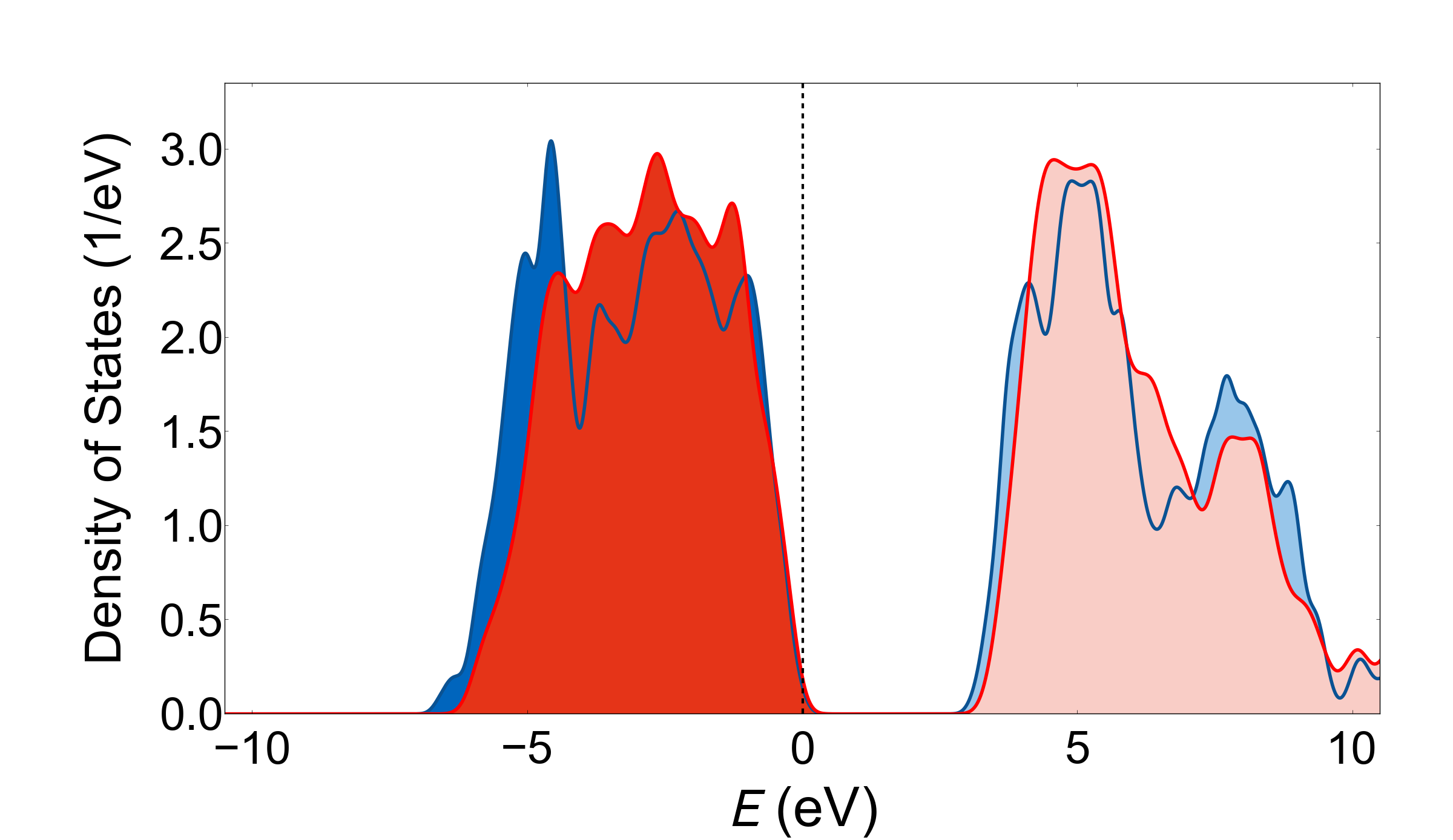}
\caption{Density of states (DOS) of the ideal TiO$_2$(110) surface per TiO$_2$ formula unit as calculated with the embedded Ti$_{46}$O$_{92}$ cluster (red) and with a supercell geometry (blue). $\epsilon_{\rm VBM}$ is used as zero reference and filled states are depicted in darker color. A Gaussian smearing ($\sigma=0.1$\,eV) is applied.}
\label{fig3}
\end{figure}

We begin our investigation by demonstrating the high-quality representation of the surface electronic structure obtained with our QM/MM setup, which could generally not be achieved without the consistent re-parametrization of the MM potentials. Figure \ref{fig3} compares the density of states (DOS) for the ideal TiO$_2$(110) surface obtained with the embedded Ti$_{46}$O$_{92}$ cluster against the results from the slab reference calculation. Good agreement is achieved for both filled and empty states. This agreement extends not only to the band gap $\Delta \epsilon_{\rm gap}$ or to the width of the valence band. Both are reproduced within 0.5\,eV and 0.9\,eV, respectively. Also more subtle features within the bands are reflected rather well. Equivalent findings are obtained for the neutral defect, where a straightforward comparison to supercell calculations is also possible (not shown).

For both the ideal surface and the neutral defect case the DOS calculated with the smaller QM clusters does also not differ much with respect to the corresponding calculation with the largest cluster. Intra-band features change, but the band gap or the width of the valence band varies each time by less than 0.2\,eV. Even on absolute scale the spectra of the smallest and largest cluster match almost perfectly. In contrast, the intermediate Ti$_{32}$O$_{64}$ cluster reveals a potential offset by 0.4\,eV up in energy, which can be attributed to its prolate shape. This cluster exposes the largest surface to volume ratio to the QM/MM interface, and is hence most affected by imperfect QM/MM coupling. Note that this potential offset is compensated in eq.~(\ref{formation}) through the correction procedure described in the previous chapter and, thus, does not affect the calculated formation energies.

Introducing the oxygen vacancy gives rise to electronic defect states.\cite{GandugliaPirovano2007219} Depending on the charge state of the defect these states are empty, singly or fully occupied. The two electrons in the fully occupied state of the V$^0_{\rm O}$ defect can thereby form three different electronic configurations with differing spin multiplicities: a closed-shell singlet, an open-shell singlet and an open-shell triplet. The closed-shell singlet corresponds to the paired excess electrons occupying the same orbital localized at the defect site. In the unpaired open-shell configurations both electrons occupy different orbitals of either parallel (triplet) or anti-parallel (singlet) spin. Rather than as intrinsic neutral V$^0_{\rm O}$ defect the latter configurations should thus be seen as a singly- (doubly-) charged defect with one (two) trapped small polaron(s).~\cite{PSSR:PSSR201206464} In the limit of large electron separation both open-shell spin configurations must become energetically degenerate. For closer distances, the different electronic configurations give rise to different formation energies. In order to determine the appropriate neutral defect state energetics for comparison to the charged states, it is therefore necessary to identify the corresponding lowest-energy electronic configuration.

As extensively discussed by Deskins {\em et al.} the polarons can be localized in numerous different patterns around the defect site with varying relative stabilities.\cite{doi:10.1021/jp2001139} Applying the same initialization tricks as applied by Deskins {\em et al.} to achieve these different localizations, we always end up with one electron trapped at the defect site, while the other electron can be trapped in different locations. Whenever these final sites are separated by more than 3\,{\AA} the two different open-shell spin configurations emerge as energetically degenerate -- exactly as reported by Deskins {\em et al.} \cite{doi:10.1021/jp2001139}. In contrast to their observation convergence to an open-shell singlet configuration proved essentially impossible for smaller electron separations in our QM/MM setup. Even when starting from geometries optimized in the triplet state and with both excess electrons initially separated, the relaxed singlet geometries always converged to the closed-shell configuration with paired electrons. Open-shell singlet states could only be stabilized in the largest QM cluster for rather large separations, then yielding energetically essentially degenerate stabilities to the open-shell triplet configuration as stated before.

\begin{figure}
\centering
\includegraphics[width=8.5 cm]{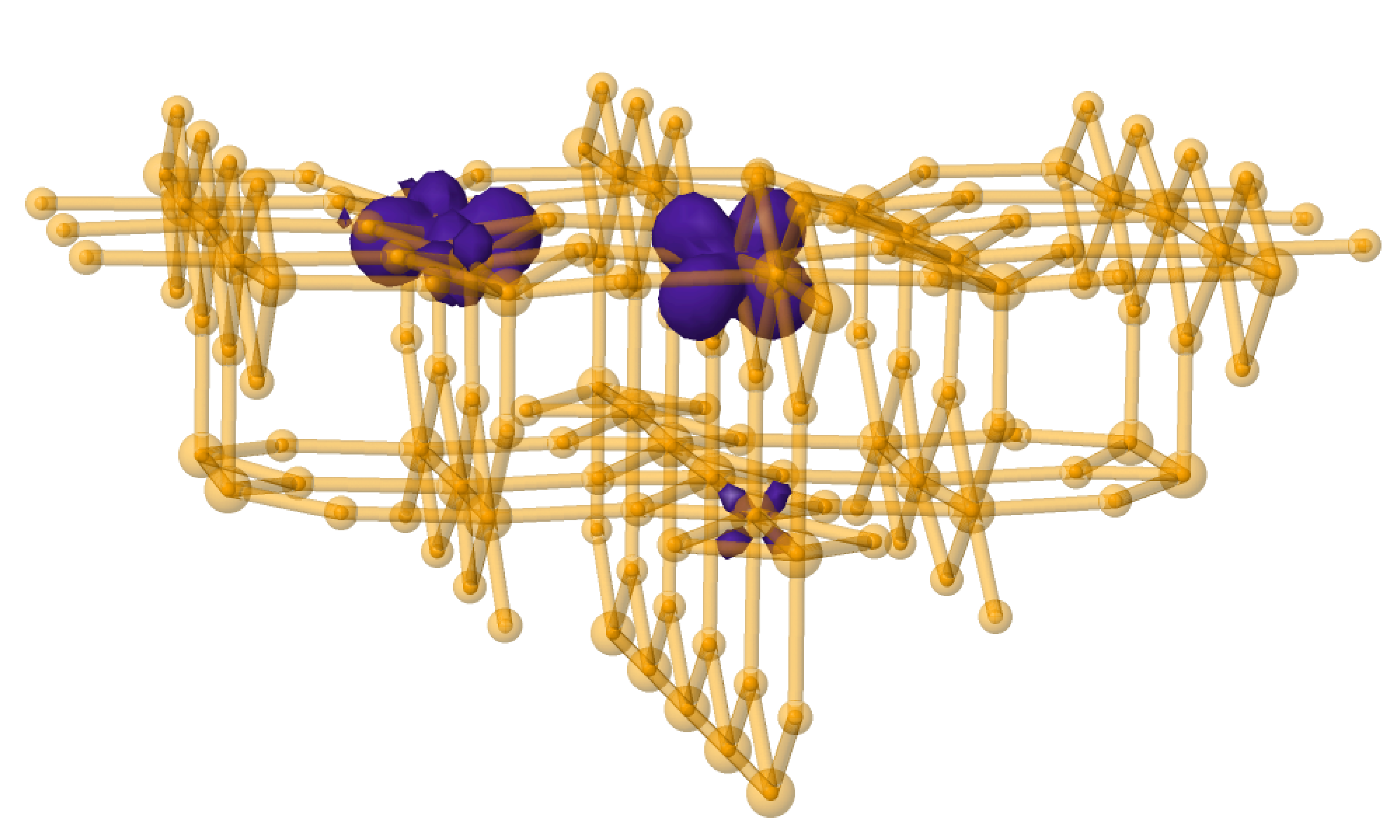}
\caption{Perspective view of the electron localization in the lowest-energy open-shell triplet conformation found for the neutral defect state with the embedded Ti$_{46}$O$_{91}$ cluster. Visualized is the electron density contour at 0.01 e/\AA$^3$.}
\label{fig4}
\end{figure}

As further discussed below the closed-shell singlet configuration is energetically much less stable than any of the various open-shell localizations. In our case the latter all exhibit rather similar relative stabilities within 0.1-0.2\,eV. This is in disagreement to Deskins {\em et al.}\cite{doi:10.1021/jp2001139} and others \cite{0953-8984-24-43-435504,PhysRevLett.105.146405,Morgan20075034,PhysRevLett.113.086402}, who reported relative stabilities varying over a range of 1.76\,eV and with the lowest-energy conformations localizing both excess electrons in the sub-surface layer. Although in principle possible in the largest Ti$_{46}$O$_{91}$ QM cluster, we can not stabilize such conformations. In contrast, the lowest-energy conformation found consistently across all three cluster sizes is depicted in Fig.~\ref{fig4}, showing one polaron localized at the defect Ti-sites of the bridging row and one on a neighboring five-fold coordinated surface Ti-atom with very little contributions at sub-surface Ti-sites. This difference to the preceding studies \cite{doi:10.1021/jp2001139,0953-8984-24-43-435504,PhysRevLett.105.146405,Morgan20075034,PhysRevLett.113.086402} could potentially arise out of remaining finite-size effects of our QM/MM setup at even the largest QM cluster employed. Alternatively, it could be due to their different treatment of the electronic structure at the effective GGA+U level. In this respect, we note that the surface localization site of our most stable conformation is in perfect agreement with previous results by Di Valentin and co-workers employing the B3LYP hybrid functional.~\cite{Selloni2006} 

Having potentially missed the true lowest-energy solution puts a certain uncertainty on the energetics of the open-shell neutral defect. However, as we will further discuss below, even a higher stability by 0.87\,eV that would then correspond to the sub-surface solution as reported by Deskins {\em et al.} would not critically affect our conclusions on its relative stability with respect to the charged defects states. Even though therewith of no concern to the present study we nevertheless plan to further investigate this issue with large supercell hybrid functional calculations and GGA+U QM/MM calculations. We note that an important aspect are certainly the used lattice parameters, which in many GGA+U studies are simply those of the underlying GGA functional. We find the use of incorrect lattice parameters to critically affect the electronic level positions and therewith the polaron stabilities. Relying on the hybrid-functional optimized lattice parameters the electronic defect levels obtained in our QM/MM study are highly consistent with those of previous bulk and surface hybrid-functional studies. For all three QM clusters the closed-shell singlet state is located 0.8\,eV below the CBM in the unrelaxed structure, and shifts up by 0.2\,eV in energy upon including lattice relaxation, cf.~refs.~\onlinecite{janotti2010hybrid, Selloni2006}. The lowest-energy triplet solution exhibits two defect states 1.7\,eV and 0.6\,eV below the CBM in the unrelaxed structure. Upon lattice relaxation the electrons condense to two small polarons, the states of which at 1.1 and 0.9\,eV below the CBM are fully consistent with experiment\cite{Aiura19941215,PhysRevB.70.045415,Wendt27062008,SIA:SIA2013,minato2014atomic,PhysRevLett.113.086402} ({\em vide infra}).

\subsection{Formation energies of unrelaxed vacancies}

\begin{table} [t]
\centering
\begin{tabular}{l|cccc}
\hline \hline
                   & \multicolumn{4}{c}{Self-consistent MM electronic polarization}                       \\
                   & Ti$_{22}$O$_{43}$ & Ti$_{32}$O$_{63}$ & Ti$_{46}$O$_{91}$ & Slab                  \\ \hline
V$^0_{\rm O}$ singlet&    5.60             &      5.66             &     5.60              &   5.59    \\
V$^{+}_{\rm O}$   &      3.19             &      3.22             &     3.17              &           \\
V$^{2+}_{\rm O}$   &      3.98             &      3.95             &     3.95              &           \\ \hline \hline
                   & \multicolumn{4}{c}{Analytic correction for electronic polarization}                  \\
                   & Ti$_{22}$O$_{43}$ & Ti$_{32}$O$_{63}$ & Ti$_{46}$O$_{91}$ &                       \\ \hline
V$^0_{\rm O}$ singlet&    5.61             &      5.66             &     5.61              &   5.59    \\                   
V$^{+}_{\rm O}$   &      3.19             &      3.21             &     3.17              &           \\
V$^{2+}_{\rm O}$   &      3.98             &      3.95             &     3.95              &           \\ \hline \hline
\end{tabular}
\caption{Unrelaxed surface defect formation energies (in eV) in the oxygen-rich limit and for a Fermi-level position at the VBM. Shown are results for the three different QM clusters employed, and, in the case of the neutral defect, also from the PBC supercell reference calculations. The data for the neutral V$^0_{\rm O}$ defect corresponds to the closed-shell singlet electronic configuration, which is the only one that can be stabilized in the absence of lattice relaxation. Upper rows correspond to a QM/MM setup, in which the electronic polarization response to the defect is treated through self-consistent relaxation of the MM shells. Lower rows correspond to a QM/MM setup, where this response is approximately obtained through an analytical polarization correction approach (see text).}
\label{table3}
\end{table}

In particular the large static dielectric response of TiO$_2$ is expected to significantly stabilize charged defects.\cite{janotti2010hybrid} In order to quantify this effect it is useful to first analyze the defect formation energies when structural relaxation in response to the defect is suppressed. From a methodological point of view, this additionally provides the possibility to assess an analytical scheme that may allow to compensate the error introduced by the inability of both the polarizable interatomic potential in QM/MM calculations and the hybrid DFT functional in PBC supercell calculations to properly account for the large static TiO$_2$ dielectric constants, cf.~Table \ref{table1}. Correspondingly, Table \ref{table3} summarizes calculated formation energies, in which thus only the high-frequency, electronic TiO$_2$ polarization is included that can accurately be accounted for by both the polarizable interatomic potential and the hybrid functional. Shown is data for $\Delta E^{\rm f}$ in the O-rich limit and for the Fermi-level position at the VBM. The series of three QM cluster sizes demonstrates a fast convergence. For the neutral defect the obtained formation energy agrees furthermore perfectly with the corresponding value obtained by a PBC supercell calculation.

In these QM/MM calculations the full electronic polarization response of the material outside the QM region is captured through the self-consistent relaxation of the O shells within the MM region, while the analytical correction $\Delta E_{\rm pol}(q)$ in eq.~(\ref{eq:pol_surface}) additionally accounts for the long-range contribution outside of the MM region. In these calculations the $\epsilon$ value entering $\Delta E_{\rm pol}(q)$ corresponds -- as appropriate for the unrelaxed case -- to the isotropically averaged high-frequency dielectric constants, cf.~Section IIC. Table \ref{table3} also compiles results obtained with a more approximate method, where the entire electronic polarization response of the material outside of the QM region is accounted for through the analytic correction equation, eq.~(\ref{eq:pol_surface}). Here, the O shells in the MM region are no longer allowed to relax after the creation of the defect. Instead, the analytical correction equation is employed with a radius $R$ that corresponds to the outer radius of the QM region instead of the outer radius of the MM region used before. This method is numerically more advantageous as it does not need to achieve self-consistency between the MM shell polarization and the QM cluster. It also provides an elegant way to later on assess the error introduced by the erroneous description of the static dielectric constants through the polarizable force field. In principle, any value for $\epsilon$ can be employed in $\Delta E_{\rm pol}(q)$, i.e.~also the "correct" experimental value. On the other hand, the method is more approximate and phenomenological. The polarization is only described at the isotropic continuum level and the outer boundary of the finite QM clusters is less well approximated by a hemisphere as the outer boundary of the much larger MM region. The results thus also depend more sensitively on the exact choice of the QM cluster radius $R$ employed in $\Delta E_{\rm pol}(q)$. For the data compiled in Table \ref{table3} the $R$ for each cluster size has been chosen as a fit parameter to reproduce the corresponding self-consistent polarization results for both charged defects. The values obtained for $R$ are for each cluster about 0.7-0.9\,{\AA} larger than what would be obtained from the positions of the outermost QM atoms, cf.~Section IIC, thus effectively accounting for the somewhat larger extension of the electron density. Intriguingly, for each cluster size the same value of $R$ can accurately reproduce the formation energies of both charged defects, cf.~Table \ref{table3}. This shows that this approximate method can be applied without further systematic errors and that for the polarization response outside the QM region any electrostatic multipole moment of the electron density higher than the monopole term considered in $\Delta E_{\rm pol}(q)$ can be indeed neglected.

The data in Table \ref{table3} shows that already the comparatively small electronic polarization is sufficient to largely stabilize the charged defects against the neutral one. We can quantify this stabilization through the calculated value of $\Delta E_{\rm pol}(q)$ in case of the approximate method, i.e.~when this term accounts for the entire response outside of the QM region. Even for the largest QM cluster, where this additional stabilization through the far-range response is smallest, $\Delta E_{\rm pol}(q)$ is still $-0.60$\,eV for the singly-charged V$^{+}_{\rm O}$ defect, whereas for the doubly-charged V$^{2+}_{\rm O}$ defect it even rises to $-2.40$\,eV. For the smallest QM cluster the corresponding values are $-0.79$\,eV and $-3.16$\,eV, respectively. For the neutral V$^{0}_{\rm O}$ defect $\Delta E_{\rm pol}(q=0) = 0$. The essentially identical values for $\Delta E^{\rm f}(V^0_{\rm O})$ obtained with the self-consistent polarization and with the analytical correction method in Table \ref{table3} thus reveal that electronic stabilization outside the QM region is in this case negligible already for the smallest QM cluster. 

This different polarization response for neutral and charged defects is sufficient to make the latter thermodynamically more stable for a wide range of Fermi-level positions. The value of $\epsilon_{\rm F}$ where the formation energies of charge state $q$ and $q'$ become equal define the transition level $\epsilon(q/q')$. With the values of Table \ref{table3} and eq.~(\ref{formation}) we determine the transition level $\epsilon(+/0)$ as 2.44\,eV, i.e.~less than 1\,eV below the CBM. Already when only accounting for the (small) electronic response, the neutral defect level on which preceding calculations largely focused their attention is therefore not stable for Fermi-level positions over a large part of the band gap. When next also considering the much larger lattice polarization of TiO$_2$, this trend will be significantly enhanced, further disfavoring the neutral defect.

\subsection{Formation energies of relaxed vacancies}

\begin{table} [t]
\centering
\begin{tabular}{l|ccc}
\hline \hline
                       & \multicolumn{3}{c}{Self-consistent MM electronic polarization}       \\
                       &  Ti$_{22}$O$_{43}$ & \hspace{0.7cm}Ti$_{32}$O$_{63}$ & Ti$_{46}$O$_{91}$      \\ \hline
V$^0_{\rm O}$ singlet  &       5.40       &\hspace{0.7cm}    5.50              &     5.33           \\
V$^0_{\rm O}$ triplet  &        4.89       & \hspace{0.7cm}    4.78              &     4.83           \\
V$^{+}_{\rm O}$       &       2.02       & \hspace{0.7cm}    1.83              &     1.92           \\
V$^{2+}_{\rm O}$       &       0.43       & \hspace{0.7cm}   -0.20              &    -0.37           \\ \hline \hline
                       & \multicolumn{3}{c}{Analytic correction for full polarization}  \\
                       & Ti$_{22}$O$_{43}$ & \hspace{0.7cm}Ti$_{32}$O$_{63}$ & Ti$_{46}$O$_{91}$  \\ \hline
V$^0_{\rm O}$ singlet  &        5.40       & \hspace{0.7cm}    5.50              &     5.33           \\
V$^0_{\rm O}$ triplet  &        4.89       & \hspace{0.7cm}    4.78              &     4.83           \\
V$^{+}_{\rm O}$       &        1.70       & \hspace{0.7cm}    1.56              &     1.68           \\
V$^{2+}_{\rm O}$       &       -0.69       & \hspace{0.7cm}   -1.27              &    -1.30           \\ \hline \hline
\end{tabular}
\caption{Relaxed surface defect formation energies (in eV) for the three different QM clusters in the O-rich limit and for a Fermi-level position at the VBM. Upper rows correspond to a QM/MM setup, in which only the electronic polarization response outside of the QM region is accounted for through self-consistent relaxation of the MM shells. Lower rows correspond to a QM/MM setup, where the full lattice and electronic response is approximately obtained through an analytical polarization correction approach (see text).}
\label{table4}
\end{table}

The results of the preceding section already indicate the relevance of the charged defect states, and correspondingly the hitherto barely explored necessity to reliably describe them with electronic structure calculations. When moving to the fully relaxed defect formation energies, our QM/MM scheme is limited by the shortcomings of the polarizable interatomic potential with respect to the static dielectric constants. Thus restricting structural relaxation to the QM region (and maintaining the full electronic polarization as in the previous section), we obtain the defect formation energies compiled in Table \ref{table4}. Comparing the entries for the neutral defect in Table \ref{table3} and \ref{table4} we observe only a minute lowering of $\Delta E^{\rm f}$ by 0.2\,eV for the closed-shell singlet electronic configuration (henceforth denoted as singlet) at all three cluster sizes. The additional lattice response accounted for in the relaxed calculations is thus quickly converged over the few atomic shells contained in the finite QM clusters. While only a marginal quantitative effect for the singlet, it is only this short-range structural relaxation that stabilizes the neutral open-shell singlet and triplet configurations at all. Nevertheless, also for the degenerate lowest-energy such configurations (henceforth denoted as triplet/singlet) quick convergence with the size of the QM region is achieved.

In line with the expectations from the large static TiO$_2$ dielectric constants, the additional lattice-polarization stabilization of the charged defects is instead significant even when only accounting for the relaxation of the nearest-neighbor atomic shells in the QM region. Comparing again the corresponding entries in Tables \ref{table3} and \ref{table4}, this stabilization is of the order of 2\,eV for the singly-charged defect and amounts even to around 4\,eV for the doubly-charged defect. As polarization of the environment scales with the square of the charge, cf.~eq.~(\ref{eq:pol_surface}), this drastic increase is not unexpected. At the latest for the $V^{2+}_{\rm O}$ defect no convergence can concomitantly be achieved over the present range of QM cluster sizes. In view of the strong lattice relaxations calculated for this defect, this is also not surprising. With an outwards relaxation of up to 0.45\,{\AA} the nearest-neighbor Ti atoms around the defect exhibit the largest displacements. Corresponding displacements in the two larger Ti$_{32}$O$_{63}$ and Ti$_{46}$O$_{91}$ QM clusters differ by less than 0.05\,{\AA}, proving that this shortest-range polarization contribution is well converged at these cluster sizes. However, in the largest Ti$_{46}$O$_{91}$ QM cluster the next two shells of Ti atoms still show maximum displacements of 0.12\,{\AA} and 0.05\,{\AA}, respectively. These non-negligible relaxation contributions can no longer be captured with the smaller clusters.

Just accounting for the lattice relaxation inside the finite QM region is therefore not sufficient to reliably determine the formation energies of the charged defects. The relaxation patterns possible inside the largest Ti$_{46}$O$_{91}$ QM cluster would require a $c(2 \times 5)$ surface unit-cell in PBC supercell calculations. Due to the periodic images the relaxation patterns outside this unit cell are spurious. Our findings for the QM/MM setup are therefore paralleled by the equivalent insight that even correspondingly large surface unit-cells are not sufficient to absolutely converge charged defects in PBC calculations. As these system sizes are at the upper limit of what is presently tractable at the hybrid functional level, extrapolation procedures are thus required within the PBC approach.~\cite{janotti2010hybrid} Within our QM/MM approach we can find an analogue in the analytic correction procedure described in the last section. Indeed, using exactly the same radius $R$ for each of the three cluster sizes as established in the unrelaxed calculations we can reproduce the $\Delta E^{\rm f}$ of Table \ref{table4} with the same accuracy as was the case in Table \ref{table3} (not shown). Here, this means that we used for $\epsilon$ in $\Delta E_{\rm pol}(q)$ of eq.~(\ref{eq:pol_surface}) exactly the high-frequency value that corresponds on average to the ones of the polarizable interatomic potential.

With this confidence we apply the approximate analytic correction also for the calculation of fully relaxed formation energies. For this, we now use as $\epsilon$ in $\Delta E_{\rm pol}(q)$ the isotropic average of the large static dielectric constants derived from HSE06, cf.~Table \ref{table1}. Keeping exactly the same radii $R$ as before in the calculation of $\Delta E_{\rm pol}(q)$, this approach thus accounts for the full lattice and electronic relaxation explicitly in the QM region, and accounts for the same full lattice and electronic polarization on the continuum level outside of the QM region. The results compiled in Table \ref{table4} indicate another sizable stabilization in particular for the doubly-charged $V^{2+}_{\rm O}$ defect, the formation energy of which is now also converged over the two larger QM clusters. Likely, the obtained $\Delta E^{\rm f}$ values nevertheless still slightly overestimate the true relaxed formation energies, as the outermost atoms in the QM region are not allowed to relax to avoid artifacts from an imperfect QM/MM coupling. From small variations of the employed radius $R$ in $\Delta E_{\rm pol}(q)$ to also effectively account for the corresponding shell of atoms, we estimate that this relaxation would lower the formation energy of $V^{+}_{\rm O}$ ($V^{2+}_{\rm O}$) by another 0.1\,eV (0.3\,eV). 

We therefore arrive at final values for the formation energies in the O-rich limit and for a Fermi-level position at the VBM of $5.3\pm0.1$\,eV (V$^0_{\rm O}$ singlet), $4.8\pm0.1$\,eV (V$^0_{\rm O}$ triplet/singlet), $1.6\pm0.2$\,eV (V$^{+}_{\rm O}$) and $-1.6\pm0.3$\,eV (V$^{2+}_{\rm O}$). The stated error bars hereby reflect conservative estimates accounting for the uncertainties implied by the analytic correction approach and the convergence with the QM cluster size. This uncertainty in particular for the doubly-charged vacancy is not entirely satisfying. Notwithstanding, we note that a similar uncertainty did arise in the calculation of the bulk defect formation energies through the extrapolation procedure required in the employed PBC supercell approach \cite{janotti2010hybrid}, whereas a reliable polarization-converged calculation of $\Delta E^{\rm f}$ of the charged surface defects at the hybrid-functional level has never even been attempted. Intriguingly, our results are on the contrary very robust against the inaccuracy introduced by the still large deviation of the static HSE06 dielectric constants with respect to experiment, cf.~Table~\ref{table1}. As apparent from the nature of eq.~(\ref{eq:pol_surface}) values for $\epsilon$ smaller than the HSE06-derived value will cause large variations in the long-range polarization correction. Varying the $\epsilon$ employed in the analytic correction from the small value corresponding to the MM high-frequency dielectric constants to the one corresponding to the static HSE06 dielectric constants thus led to the just discussed large changes of the formation energies of the charged defects. However, a further increase of $\epsilon$ to match the experimental dielectric properties of Table~\ref{table1} instead yields only a negligible further stabilization of the singly- (doubly-) charged defect by 0.01\,eV (0.03\,eV).

In total, structural relaxation therefore lowers the formation energies of the closed-shell $V^0_{\rm O}$ singlet by 0.3\,eV, of V$^{+}_{\rm O}$ by 1.6\,eV, and of V$^{2+}_{\rm O}$ by 5.6\,eV. In the work by Janotti {\em et al.} for the bulk O vacancy the corresponding values were 0.3\,eV, 0.9\,eV and 3.5\,eV.~\cite{janotti2010hybrid} This suggests a significantly stronger stabilization of the charged defects at the surface, which one may attribute to a generally larger structural flexibility of surface atoms. In light of the preceding discussion on the insensitivity of the polarization-response to small variations away from the HSE06 dielectric properties, this conclusion should not be affected by the use of a tailored HSE functional with a slightly different exact exchange mixing in this preceding work. More likely, the different functional will affect the description of the short-range QM rehybridization and therewith prevent a direct comparison of the absolute defect formation energies in the bulk and at the surface. Compared to our surface values of 5.3\,eV (V$^0_{\rm O}$ singlet), 1.6\,eV (V$^{+}_{\rm O}$) and $-1.6$,eV (V$^{2+}_{\rm O}$), Janotti {\em et al.} report 5.3\,eV, 2.2\,eV and $-1.4$\,eV for the bulk, respectively.~\cite{janotti2010hybrid} Overall, the numbers are intriguingly similar. Considering also the uncertainties due to PBC extrapolation or QM/MM convergence we can, however, not make any more detailed statements. Even though the calculated $\Delta E^{\rm f}({\rm V}^{2+}_{\rm O})$ reflect the same trend we can therefore not comment on the strong surface segregation tendency obtained previously at the GGA or GGA+U level.~\cite{doi:10.1021/jp811288n,Bjorheim20128110} Corresponding (polarization-response unconverged) calculations had predicted a higher stability of the surface O vacancy by $\sim 0.7-1$\,eV as compared to its counterpart in the bulk.

\subsection{Thermodynamic stability}

 \begin{figure}[h]
 \centering
 \includegraphics[width=7 cm]{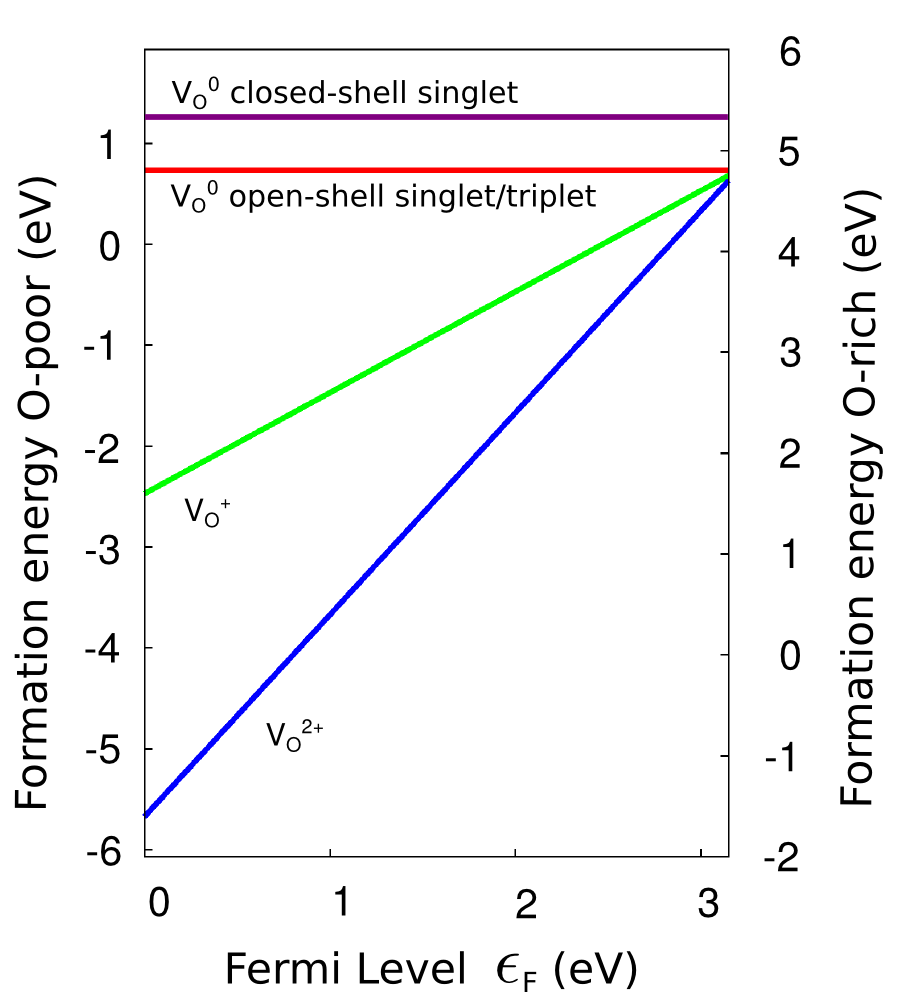}
 \caption{Calculated relaxed surface defect formation energies as a function of the Fermi-level position in the band gap. $\epsilon_{\rm VBM}$ is used as zero reference for $\epsilon_{\rm F}$. The y-axis scales on the left and right correspond to $\Delta E^{\rm f}$ in the the O-poor and O-rich limit, respectively, and are correspondingly shifted by 4.07\,eV, cf.~eq.~(\ref{O-poor}).}
 \label{fig5}
 \end{figure}

As clear from eq.~(\ref{formation}) the defect formation energies depend on the reservoirs available for the removed species, i.e.~the oxygen chemical potential and the electronic Fermi level. Figure~\ref{fig5} therefore extends the previous discussion for the O-rich limit and for a Fermi-level position at the VBM, and shows the different $\Delta E^{\rm f}$ as a function of the Fermi-level position and with $y$-axis scales for both the O-rich and O-poor limit. For any Fermi-level position within the band gap, the doubly-charged V$^{2+}_{\rm O}$ defect exhibits the lowest formation energies; a situation that was equivalently obtained before for the bulk O vacancy.~\cite{janotti2010hybrid} When also accounting for the full lattice relaxation, the transition levels $\epsilon(2+/+)$ and $\epsilon(2+/0)$ are thus located above the CBM, i.e.~the surface O vacancy is a shallow donor.

Even in the limit of strong $n$-type doping with a concomitant Fermi level close to the CBM the energetic gap to the neutral closed-shell singlet $V^0_{\rm O}$ with two electrons bound in the localized defect orbitals is still quite large. In full agreement with previous theoretical studies \cite{doi:10.1021/jp2001139,0953-8984-24-43-435504,PhysRevLett.105.146405,Morgan20075034,PhysRevLett.113.086402,Selloni2006} the equally charge-neutral open-shell triplet/singlet is obtained as an energetically preferred electronic configuration. However, in contrast to the intrinsically neutral defect, this would instead correspond to a situation where the charged defect has trapped two small polarons \cite{janotti2010hybrid,PSSR:PSSR201206464}. In our calculations, this situation only becomes energetically degenerate to the bare doubly-charged V$^{2+}_{\rm O}$ defect for a Fermi-level position right at the CBM, cf.~Fig.~\ref{fig5}. Here, we have to recall though that Deskins {\em et al.} reported a polaron localization in the sub-surface layer, which we could not confirm with the present hybrid-functional QM/MM setup ({\em vide supra}). In their scheme, this sub-surface configuration was 0.87\,eV more stable than the surface polaron triplet/singlet configuration we obtain as most stable, cf.~Fig.~\ref{fig4}. If the sub-surface solution by Deskins {\em et al.} is indeed physical, this would thus lower the open-shell triplet/singlet line in Fig.~\ref{fig5}. The resulting lowering of the transition level $\epsilon(2+/0)$ to 0.44\,eV below the CBM would then indicate the possibility of polaron trapping at the surface defects for corresponding Fermi-level positions. As discussed by Janotti {\em et al.} \cite{PSSR:PSSR201206464} it is such trapped polarons (trapped possibly at the surface O defect but equally at other surface and bulk defects), not the O surface defect itself that are responsible for the defect state in the band gap observed in numerous experimental studies \cite{GandugliaPirovano2007219}.

For Fermi-level positions further away from the CBM, the formation energy of the V$^{2+}_{\rm O}$ defect decreases rapidly, cf.~Fig.~\ref{fig5}. Under O-rich conditions it is only this steep lowering of $\Delta E^{\rm f}$ that eventually leads to values consistent with appreciable defect concentrations. This is consistent with experimental reports on increased defect concentrations upon $p$-doping.\cite{CHEM:CHEM200700472,doi:10.1021/cr00033a004,PhysRevLett.96.026103} Equivalent reductions of $\Delta E^{\rm f}$ (and concomitant increases of the defect concentration) can, of course, also be achieved with less O-rich conditions, i.e.~by lowering the O chemical potential. A strong presence of bridging-oxygen vacancies at the TiO$_2$(110) surface has indeed frequently been observed in ultra-high vacuum experiments.\cite{schaub2001oxygen,Diebold2003,Henderson2011185,PhysRevLett.100.055501} As apparent from Fig.~\ref{fig5} the lowering of $\Delta E^{\rm f}$ with lower $\epsilon_{\rm F}$ or $\mu_{\rm O}$ is in fact so strong, that we eventually obtain negative formation energies. This unphysical result indicating a lattice instability is an artifact of the persistent use of defect formation energies calculated for the dilute limit. Under conditions corresponding to increased defect concentrations, defect-defect interactions as well as the build-up of a space-charge region \cite{richter2013concentration} would in reality modify the formation energies to suppress negative values.\cite{richter2013concentration} In view of the huge energetic preference of the doubly-charged defect presently obtained for such conditions, it is, however, unlikely that this will change the energetic ordering of neutral and charged defects.

\section{Summary and Conclusions}

We presented a solid-state QM/MM approach with a polarizable interatomic potential optimized to match the DFT xc functional employed in the QM region and analytical corrections for long-range polarization effects. Corresponding embedded-cluster calculations provide a first determination of the defect formation energies of neutral and charged O vacancies at the TiO$_2$(110) surface at the hybrid-functional DFT level and containing a converged contribution of the strong dielectric response of this material. The stabilization of the singly- ($V^{\rm +}_{\rm O}$) and doubly- ($V^{\rm 2+}_{\rm O}$) charged surface defect through lattice relaxation is indeed found to be sizable, i.e.~of the order of 1.6\,eV and 5.6\,eV, respectively. It is thus even larger than previously obtained for the bulk O vacancy\cite{janotti2010hybrid}, a fact that we attribute to the generally larger structural flexibility of surface atoms.

The stabilization of in particular the doubly-charged $V^{\rm 2+}_{\rm O}$ defect is large enough to make it the thermodynamically most stable state for any Fermi-level position in the band gap. However, under the uncertainties of our approach we cannot exclude a possible trapping of small polarons at the charged defect for Fermi-level positions close to the CBM. The situation for the surface O vacancy would then be fully equivalent to the one discussed by Janotti {\em et al.} before for the bulk.\cite{janotti2010hybrid,PSSR:PSSR201206464} The surface O vacancy is thus a shallow donor and the electronic defect state within the band gap observed experimentally results from trapped polarons, not from the intrinsic O defect itself.

Within the nature of a charged defect, the formation energy of the $V^{\rm 2+}_{\rm O}$ surface O vacancy varies with the Fermi-level position in the band gap. In line with experimental reports this predicts largely increased vacancy concentrations upon $p$-doping. As surface vacancies are frequently discussed as reactive centers, systematic variations of doping concentrations may therefore provide an important avenue to tune the catalytic activity of TiO$_2$. The presented embedded-cluster approach allows to efficiently address charged reaction intermediates or their binding to charged defects both under a converged account of the large polarization response and at hybrid-functional DFT or even beyond. This predestines the approach to quantitatively assess this avenue for this and other (polarizable) metal oxides.\\

\section{Acknowledgements}

DB acknowledges a scholarship from the TUM International Graduate School of Science and Engineering. Further support came from the Solar Technologies Go Hybrid initiative of the State of Bavaria. The authors acknowledge the use of the following high-performance computing facilities, and associated support services: SUPERMUC, provided by the Leibniz-Rechenzentrum Garching; HYDRA, provided by the Max-Planck Supercomputing Center.

\end{document}